# Realization of Non-Hermitian Hopf Bundle Matter

Yung Kim[1,2][†], Hee Chul Park[3,4][†], Minwook Kyung[1], Kyungmin Lee[1], Jung-Wan Ryu[3], Oubo You[5], Shuang Zhang[5], Bumki Min[1][*], Moon Jip Park[3,6][*]

[1]*Department of Physics, Korea Advanced Institute of Science and Technology (KAIST), Daejeon 34141, Republic of Korea*

[2]*Department of Mechanical Engineering, Korea Advanced Institute of Science and Technology (KAIST), Daejeon 34141, Republic of Korea*

[3]*Center for Theoretical Physics of Complex Systems, Institute for Basic Science, Daejeon 34126, Republic of Korea*

[4]*Department of Physics, Pukyong National University, Busan 48513, Republic of Korea*

[5]*Department of Physics, University of Hong Kong, Hong Kong, 999077, China*

[6]*Department of Physics, Hanyang University, Seoul 04763, Republic of Korea*

[*]Correspondence to: bmin@kaist.ac.kr or moonjippark@gmail.com.

[†]These authors contributed equally to this work.

# Abstract

Line excitations in topological phases are a subject of particular interest because their mutual linking structures encode robust topological information of matter. It has been recently shown that the linking and winding of complex eigenenergy strings can classify one-dimensional non-Hermitian topological matter. However, in higher dimensions, bundles of linked strings can emerge such that every string is mutually linked with all the other strings. Interestingly, despite being an unconventional topological structure, a non-Hermitian Hopf bundle has not been experimentally clarified. Here, we make the first attempt to explore the non-Hermitian Hopf bundle by visualizing the global linking structure of spinor strings in the momentum space of a two-dimensional electric circuit. By exploiting the flexibility of reconfigurable couplings between circuit nodes, we can study the non-Hermitian topological phase transition and gain insight into the intricate structure of the Hopf bundle. Furthermore, we find that the emergence of a higher-order skin effect in real space is accompanied by the linking of spinor strings in momentum space, revealing a bulk-boundary correspondence between the two domains. The proposed non-Hermitian Hopf bundle platform and visualization methodology pave the way to design new topologically robust non-Hermitian phases of matter.

Studying the phases of matter that feature string and loop excitations is an emerging area in the field of condensed matter physics [1–3]. Realizing such matters would bridge the gap between our knowledge of physics and the mathematical field of topology [4-24]. A recent interesting proposal utilizes the nature of the complex eigenenergy in non-Hermitian systems to show that one-dimensional systems described by *complex eigenenergy strings* can be classified by topological linking and winding invariants [25-32]. This non-Hermitian linking structure could therefore offer a promising approach for elucidating the robust physical quantity in the presence of open-environmental couplings [33-42]. Higher-dimensional systems, in general, tend to allow more intricate linking structures [43], but the experimental visualization and reconstruction of these linking structures are still lacking.

As a first step towards generalizing such linked-string matter, we introduce two-dimensional non-Hermitian electric circuit systems, whose Bloch Hamiltonian is described by the effective spinor representation. In situ control of the circuit network allows adiabatic deformations of the Hamiltonian (Fig. 1a). The adiabatic trajectory of equal spinors generally forms a line, which we refer to as spinor strings (SS). Such string representations of the non-Hermitian Hamiltonian are naturally characterized by the nontrivial topology that arises from the mutual linking of multiple strings. Mathematically, the bundle of SSs admits a linking structure, known as the Hopf bundle [44], in which each SS is linked to all the other strings.

While the Hopf bundle has been suggested as a tool for the topological comprehension of Hermitian systems [17,19], its influence on the boundary effects in non-Hermitian systems has yet to be determined. Here, we reveal a new type of bulk-boundary correspondence (BBC) associated with non-Hermitian Hopf bundles. It is shown that the nontrivial deformation of the Hamiltonian or the Hopf bundle, rather than the Hamiltonian itself, gives rise to the boundary states (or corner skin states). We demonstrate this in this work by employing a non-Hermitian electric circuit platform capable of exhibiting various Hopf bundle projections. Realizing such condensed phases of many linked strings would allow for the discovery of a new type of intricate non-Hermitian topological continuum.

**Results**

**Non-Hermitian topological phases and Hopf bundle visualization**

We consider the generic deformable Hamiltonian of a two-dimensional non-Hermitian lattice consisting of unit cells with two sublattice sites,

$$H(\mathbf{k}, \phi) = if_0(\mathbf{k}, \phi)I_2 + f_x(\mathbf{k}, \phi)\sigma_x + f_y(\mathbf{k}, \phi)\sigma_y + f_z(\mathbf{k}, \phi)\sigma_z, \quad (1)$$

where $\sigma$ represents the Pauli matrices for the sublattice degree of freedom and $\phi \in [0, 2\pi)$ is the periodic deformation parameter. The effective spinor $\mathbf{z}(\mathbf{k}, \phi) = \left(\bar{f}_0 + i\bar{f}_x, \bar{f}_y + i\bar{f}_z\right)^T$ parameterizes the non-Hermitian Hamiltonian in Eq. (1) by encoding both information of the eigenstates and the finite lifetime arising from the non-Hermiticity. Furthermore, its spin expectation value $\mathbf{m}(\mathbf{k}, \phi) = \mathbf{z}(\mathbf{k}, \phi)^\dagger \boldsymbol{\sigma} \mathbf{z}(\mathbf{k}, \phi)$ provides the visualization of the Hamiltonian projected from the momentum space onto the Bloch sphere ($S^2$). Under deformation, the trajectory of the spinors forms a line (SS) in the three-dimensional toric space [$(k_x, k_y, \phi) \in T^3$]. By counting the linking numbers between different SSs, the adiabatic deformations of the Hamiltonian can be topologically classified. The non-Hermitian Hopf bundle corresponds to nontrivial linking ($W = 1$), and in general, the topological Hopf invariant ($\pi_{T^3}(S^2) = \mathbb{Z}$) can be defined to count the linking number between different SSs,

$$W = \frac{1}{2\pi^2} \int \epsilon^{ijkl} \bar{f}_i d\bar{f}_j d\bar{f}_k d\bar{f}_l. \quad (2)$$

Here, $\epsilon^{ijkl}$ is the Levi–Civita symbol (see Supplementary Information B for a detailed calculation) [45].

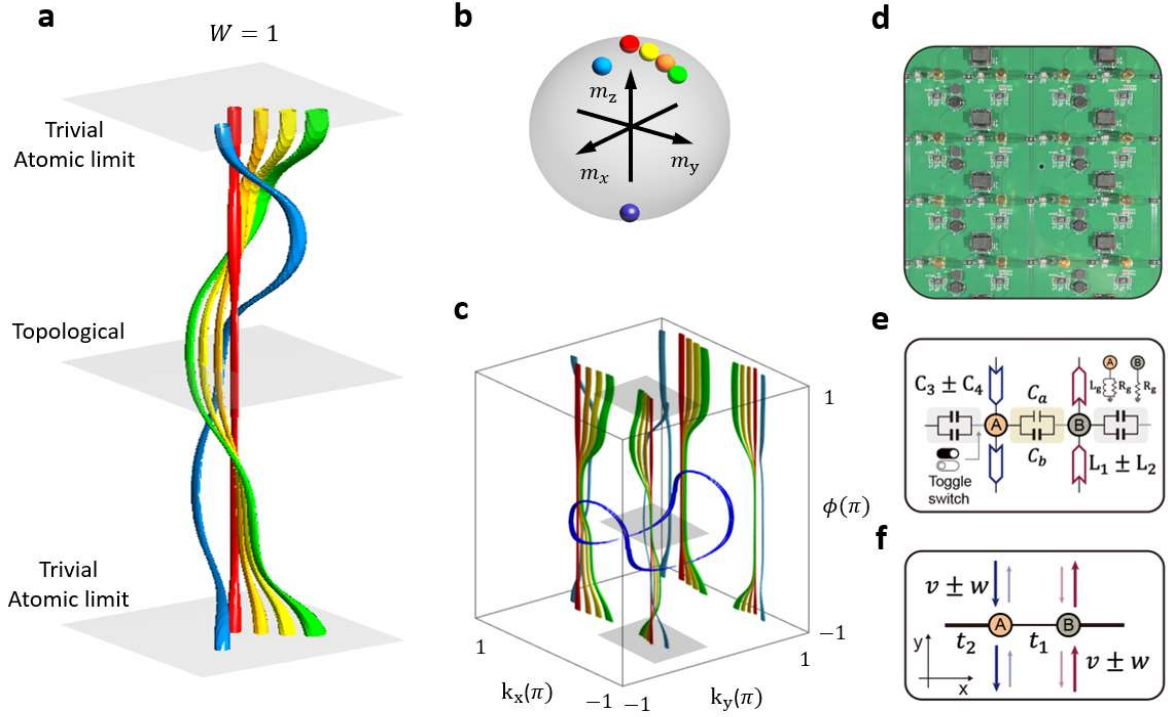

**Figure 1 Non-Hermitian Hopf bundle and realization in circuit platform.** (a) Spinor strings of the non-Hermitian Hopf bundle ($W = 1$). The nontrivial Hopf bundle is characterized by mutual links between the SSs. (b) Colour map of different spinors. Each spinor represents the corresponding Bloch Hamiltonians. (c) The spinor strings in the full Brillouin zone (Fig. 1(a) represents the shaded region). (d) Photographic image of the circuit network designed for the realization of Hopf bundle matter. (e) Layout of the unit cell consisting of circuit nodes A and B. Parallelly connected capacitors $C_a$ and $C_b$ are used to bridge the nodes, and toggle switches are used to selectively turn on/off the constituting capacitors along the effectively dimerized direction. To achieve anisotropic intercell coupling between unit cells in the $y$-direction, INICs with capacitors $C_3$ and $C_4$ and inductors $L_1$ and $L_2$ are employed to couple A-A and B-B nodes, respectively. (f) Tight-binding model of the unit cell. The red and blue arrows denote the staggered anisotropic hopping amplitudes $v \pm w$ between the unit cells along the $y$-direction. The shallow and bold lines denote intracell hopping $t_1$ and intercell hopping $t_2$, respectively.

To exemplify this concept, we design and fabricate a two-dimensional circuit network (a square lattice of 8×8) on a printed circuit board (Fig. 1d), which allows us to design the deformable Hamiltonian in the momentum space. Specifically, the circuit is designed to achieve a Su–Schrieffer–Heeger (SSH) chain along the $x$-direction while emulating a Hatano-Nelson-like chain along the $y$-direction. The dimerized couplings along the $x$-direction mimic the SSH chain with intracell ($t_1$) and intercell ($t_2$) couplings, as depicted by the thin and thick black lines in

Fig. 1f. The arrows indicate staggered directions of the nonreciprocal (asymmetric) couplings $v \pm w$ in the y-direction, with the colours of the arrows representing the sublattice sites (orange and grey for A and B sites, respectively). The Hamiltonian undergoes four characteristic configurations, depicted in Extended Data Fig. S1a [second-order topological ($H_{SO}$), first-order nonreciprocal ($H_{NR}$), first-order dimerized phases ($H_{SSH}$) and trivial atomic limit ($H_{triv}$)], as a result of gradual adjustment of dimerization and nonreciprocity along the x- and y-directions. More specifically, we consider the adiabatic deformation of the Hamiltonian,

$$H(\boldsymbol{k}, \phi) = A(\phi)H_{SO}(\boldsymbol{k}) + B(\phi)H_{NR}(\boldsymbol{k}) + C(\phi)H_{triv}(\boldsymbol{k}) + D(\phi)H_{SSH}(\boldsymbol{k}), \quad (3)$$

where $A, B, C, D$ are the Gaussian-shaped deformation parameters shown in Fig. S1b (see Supplementary Information B for more detail). When the deformation consecutively encircles the four distinct Hamiltonian phases, the Hopf invariant corresponds to $W = 1$ (i.e., the existence of the Hopf bundle, which is shown in Fig.1a). In general, the change in the Hopf invariant is accompanied by the touching of distinct SSs, where the SSs become ill-defined (i.e., vanishing of the Bloch Hamiltonian).

In the circuit implementation (Fig.1e), the two sublattices within each unit cell are represented by Circuit Nodes A and B. Along the effectively dimerized direction, the intracell and intercell couplings are made switchable by combining toggle switches in tandem with the parallelly connected capacitors $C_a$ and $C_b$. To observe the Hamiltonian deformation, we can individually vary the intracell and intercell couplings (characterized by effective capacitances $C_1$ and $C_2$) by selectively activating the relevant capacitor in the branch. The negative impedance converters through current inversion (INIC) [46, 47] are positioned in the branch between unit cells with alternating orientations to achieve the required staggered pattern of nonreciprocal couplings along the y-direction. More specifically, INICs configured with Capacitors $C_3$ and

$C_4$ are used to provide nonreciprocal coupling between the A nodes along the y-direction, while INICs with Inductors $L_1$ and $L_2$ are placed in the branch between the B nodes (see Supplementary Information A for the comprehensive circuit layout). Furthermore, an inductor $L_g$ is added to match the resonance condition by connecting Node A to the ground, and a resistor $R_g$ is employed in the measurement for impedance matching. At resonance, the (scaled) circuit Laplacian derived from the fabricated platform takes the form of the above tight-binding Hamiltonian (see Supplementary Information A),

$$(i\omega_0)^{-1}J(\mathbf{k},\omega_0) = \left(-2iC_4 \sin k_y\right)I_2 + (C_1 + C_2 \cos k_x)\sigma_x + (-C_2 \sin k_x)\sigma_y + \left(-2C_3 \cos k_y\right)\sigma_z \quad (4)$$

where $\omega_0$ is the resonance frequency of $LC$ resonators ($\omega_0 = 1/\sqrt{L_1 C_3} = 1/\sqrt{L_2 C_4}$). Notably, there is a one-to-one correspondence between the tight-binding model and the circuit lattice, which can be established by equating the coupling coefficients to the capacitances in the circuit: $t_1 = C_1, t_2 = C_2, v = -C_3, w = C_4$ (see Supplementary Information A).

Our circuit configuration features switchable couplings, which serve as an example of in situ Hamiltonian (or Laplacian) deformations. To achieve this, we set the toggle switches in such a way that the three phases shown in Fig. 2 can be accessed in a single electric lattice. Specifically, the nontrivial, critical, and trivial phases of SSH chains are prepared by setting the normalized coupling difference $\Delta t = (t_1 - t_2)/(t_1 + t_2)$ to -1, 0, and 1, respectively. By controlling the relative strengths of intracellular and intercellular couplings, we sectionally visualize the Hopf bundle with Hamiltonian deformation from the second-order topological to the first-order nonreciprocal phase (see Extended Data Fig. S1a). Experimentally measured spinor configurations agree well with the theoretical predictions of the deformations (Fig. 2**a**). More specifically, we track the trajectories of the spin-up spinors (red dots) and the spin-down spinors (blue dots). The spin-up spinors and the spin-down spinors form quadruplet distributions in the second-order topological phase (Fig.2d). As the intracell and intercell couplings are gradually matched ($\Delta t = 0$), the system approaches the first-order nonreciprocal phase. The spin-down states merge and pair-annihilate (Fig. 2c). Eventually, only the spin-up SSs are observed in the momentum space without any signature of their linking with the spin-down SSs (Fig. 2b), featuring the onset of the trivial atomic limit.

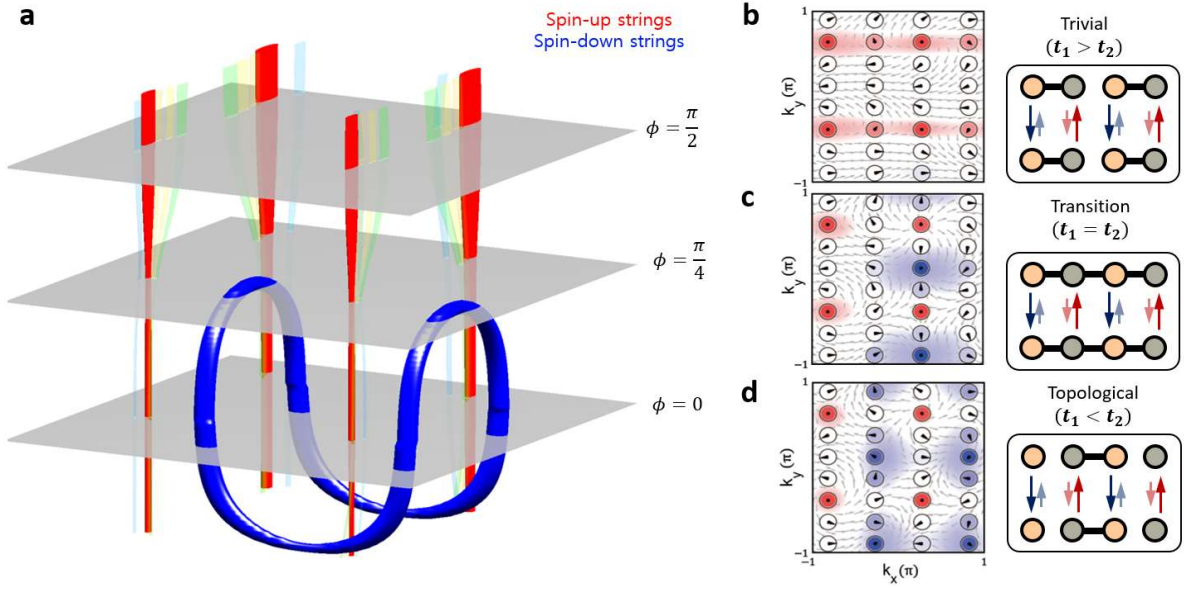

**Figure 2 Sectional visualization of the Hopf bundle**. (a) Theoretically calculated adiabatic deformation of the spinor strings. The red and blue strings represent the spin-up and the spin-down spinors, respectively. (b)-(d) Experimentally observed sectional visualization of the deformation, corresponding to $\phi = 0, \frac{\pi}{4}$ and $\frac{\pi}{2}$, respectively. The two-dimensional slice of the Hopf bundle in the first-order nonreciprocal phase shows only the spin-up SSs. The appearance points of the spin-down states start to be detached at the topological phase transition (third panel from the left). In the second-order topological phase, one of the four points of each SS is trapped inside the other four points of the other SSs.

**Bulk-boundary correspondence of non-Hermitian Hopf bundle**

We simultaneously measure complex admittance spectra and modal voltage distributions in real space (Fig. 3). To reconstruct the circuit Laplacian under the PBC and OBC, a series of current-voltage measurements are carried out with the corresponding boundary conditions by adjusting both the bulk and the boundary hopping elements. The comparison of Figs. 3**d** and **h** clearly shows the qualitative difference between the PBC and OBC spectra in the second-order topological phase ($\Delta t < 0$). Notably, we find the appearance of in-gap states (represented by red data points), which are identified as non-Hermitian skin defect states, as will be explained further below [48–59]. Since the intracell and intercell couplings are gradually matched ($\Delta t = 0$), the line gap closes ($\text{Im}[j] = 0$, see Fig. 3**c**), which coincides with the topological phase transitions between the second-order topological and first-order nonreciprocal phases. In the nonreciprocal trivial phase, the measured bulk spectrum exhibits the reoccurrence of the line gap in the complex energy plane (see Supplementary Information C for the detailed spectral analysis) [26]. Correspondingly, we observe the absorption of the in-gap states to the bulk (represented by the navy dots in the OBC spectra in Fig. 3).

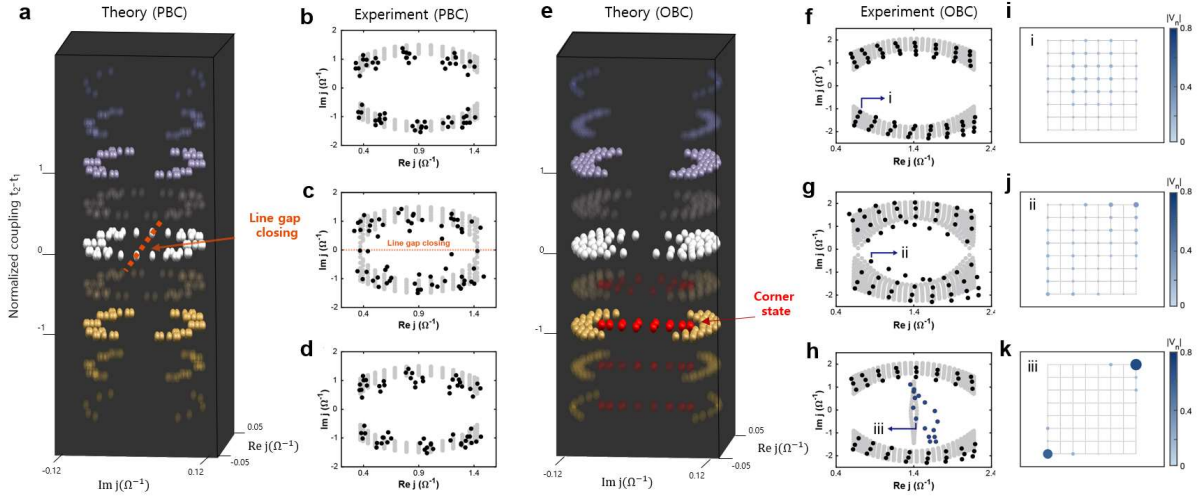

**Figure 3 Complex admittance spectra and modal voltage distributions.** Experimentally measured complex admittance spectra and modal voltage distributions under PBC and OBC. In the nonreciprocal trivial phase, only the two bulk bands are observed in the PBC and OBC spectra. At the second-order topological phase transition, the line gap of the non-Hermitian Hamiltonian closes. The experimentally measured PBC spectrum clearly shows the line gap touching the topological phase transition (c). Correspondingly, we observe the lifting of in-gap states from the bulk in the OBC spectrum (g). In the second-order topological phase, modal voltage distributions confirm 16 in-gap states within the line gap (h). The in-gap states exhibit the localized modal voltage distribution at the corner (k), which is in direct contrast with the other extended bulk states (i,j). The black dots in the complex admittance spectra denote the measured data, while the background grey dots are the calculated data. In the modal plot, the eigenvoltage amplitude is plotted with the size and colour of the circle.

The in-gap states are distinguished from the other extended bulk states by the localizations at the corner (Fig. 3k). This localization is similar to the phenomena of the second-order Hermitian topological phases [60]. However, we find that the number of localized states at the corners scale with the length of the system, which is in direct contrast with the phenomenology of the Hermitian topological phases. The anomalous scaling is indicative of the macroscopic collapse of the bulk state behaviour known as the second-order NHSE [61–65]. The emergence of the corner skin states can be naturally explained within the framework of the topological defect [26, 61, 66]. We consider the gradual variation of the system Hamiltonian in the second-order topological phase as a function of the real space position $\mathbf{r} = (R \sin \phi, R \cos \phi)$; here, $\mathbf{r}$ is defined on the loop surrounding the corner. For instance, $\phi = 0$ represents the bulk Hamiltonian inside the circuit, while $\phi = \pi$ is the vacuum Hamiltonian outside the circuit. This variation explicitly mirrors the adiabatic deformation in Extended Data Fig. S1. The existence of the topological defect within the loop manifests as the corner skin state.

The microscopic origin of the BBC can be inferred by analysing the spectral flow of the complex energy bands. Figure 4 shows the complex energy band with a nontrivial Hopf invariant plotted as a function of the spin polarization $m_z$. While two complex energy bands are separated by the imaginary gap under the PBC, their $k_y$-directional spectral flows (drawn with red arrows) exhibit opposite winding directions in the $m_z$ − Re[$j$] plane (see Fig. 4a). When the OBC is imposed in the $x$-direction (Fig.4b), two in-gap bands (highlighted by red circles) additionally appear with maximal polarizations ($m_z \approx \pm 1$). The opposite polarizations indicate that the ring states originate at the SSH boundaries and are localized at the x-directional ends. Notably, each ring state results in the counter-directional spectral flow, characterized by the opposite winding number $C(J)$ defined in the $\text{Re}[j] - \text{Im}[j]$ plane [27], i.e.,

$$C(J_0) = \frac{1}{2\pi i} \int dk_y\, \partial_{k_y} \log(j(k_y) - J_0), \quad (5)$$

where $j(k_y)$ is the admittance spectra of the ring states and $J_0 \in \mathbb{C}$ is the reference complex admittance. The nontrivial winding number, $C(J_0)$, of the ring state spectrum ensures a finite lifetime ($\propto \text{Re}[j]$) imbalance between the left ($v_g < 0$) and right moving ($v_g > 0$) states, where $v_g \propto \frac{\partial \text{Im}\, j}{\partial k_y}$ denotes the group velocity (Fig. 4c). As a result, the direction of the persistent current along the two edges of SSH chains becomes opposite. When the OBC is further imposed in the *y*-direction, the persistent current leads to macroscopic accumulations of ring states at the corner of the system.

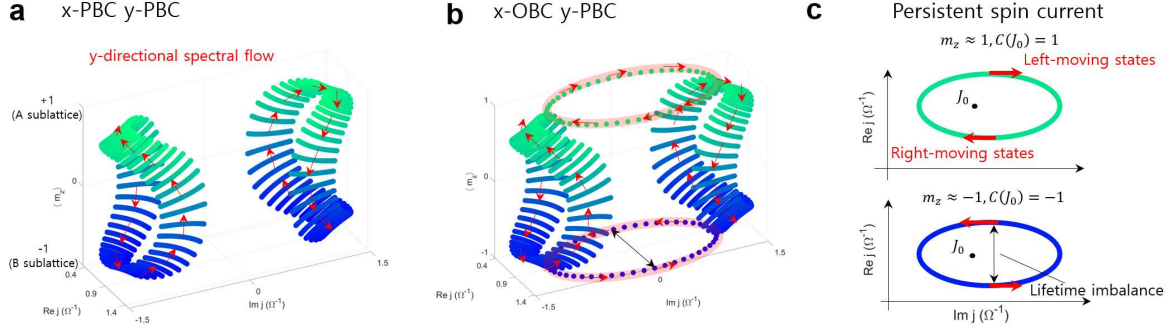

**Figure 4 Complex energy spectra and corner skin states in OBC**. Complex energy spectra of the eigenstates as a function of the spinor polarization $m_z$: (a) PBC in both the *x*- and *y*-directions, (b) OBC for the x-direction and PBC for the *y*-direction. Red arrows represent the spectral flow of the eigenstates as a function of the momentum $k_y$. (c) The spectral flow of the ring states results in a lifetime imbalance between the left- and right-moving states, which manifests as the non-Hermitian skin effect under the OBC for both the *x*- and *y*-directions.

**Discussions**

In this work, we have successfully realized a non-Hermitian Hopf bundle and mapped out the linking structure of SSs using an electric circuit lattice. The asymmetric couplings in real space are found to be crucial for achieving the non-Hermitian Hopf bundle structure in momentum space. The effective capacitance switching capability in the measurement allows three-dimensional visualization of the nontrivial linking of SSs as well as the identification of the topological phase transition by in situ Hamiltonian deformation. Using this methodology, the BBC between the non-Hermitian Hopf bundle of the bulk and the defect states at the corner is uncovered. We expect that the proposed non-Hermitian Hopf bundle platform and the visualization technique of its linking structure will serve as an intriguing testbed for the investigation and classification of non-Hermitian topology. Notably, the circuit platform enables the design of long-range nonreciprocal couplings, which may lead to the realization of Hopf bundles with higher linking numbers. In theory, new kinds of skin states would appear at the interface between two Hopf bundles with different linking numbers, wherein multiple localization length scales would coexist. Consequently, the generalization of higher-number linking structures in a non-Hermitian Hopf bundle would be an intriguing subject for future research.

**Methods**

**Sample Fabrication**

We fabricate a two-dimensional circuit network consisting of 8×8 unit cells on a printed circuit board (PCB). In the *x*-direction, we parallelly connect two different types of capacitors with a toggle switch in each branch: $C_a$ = 47 nF (WALSIN) and $C_b$ = 94 nF (KEMET). Using the toggle switch, we can choose between the capacitance values of 47 nF, 94 nF, and 141 nF. As shown in Fig. 1**e**, we use INICs to achieve asymmetric couplings in the *y*-direction; in branches connecting A nodes, operational amplifiers (Analog Devices) are configured with capacitors of $C_3$ = 47 nF and $C_4$ = 94 nF (WALSIN), while those in branches connecting B nodes contain inductors of $L_1$ = 150 μH (BOURNS) and $L_2$ = 75 μH (Abracon). In addition, a capacitor with a value of $C_s$ = 4700 nF and a resistor with a value of $R_s$ = 20 Ω are incorporated in the INICs to stabilize the circuit. To satisfy the resonance condition of the LC resonator, Node A is grounded by an inductor of $L_g$ = 18.7 μH. For impedance matching, each node is additionally grounded by a resistor $R_g$ = 20 Ω. At each node, 50-Ω SMA connectors are installed for voltage measurement.

**Experimental implementation**

The circuit Laplacian can be reconstructed by measuring the input-current and output-voltage response [67]. The measurement is conducted at the resonance frequency of $f = 60$ kHz. The AC current is inputted into the circuit by the function generator (GWinstek). We then measured the response at all nodes by using data acquisition (DAQ, National Instruments). Circuit Laplacian reconstruction can be accomplished through a series of measurement protocols with appropriate boundary conditions [46]. Under the OBC, we feed an input current at a specific node and measure the output voltage at all the other nodes. This procedure continues until all nodes in the circuit are set as an input. Under the PBC, we feed an input current at specific sublattices A and B and measure the corresponding voltage responses.

**Acknowledgments**

This work was supported by National Research Foundation of Korea (NRF) through the government of Korea (NRF- 2022R1A2C301335311). This work was also supported by the center for Advanced Meta-Materials (CAMM) funded by Korea Government (MSIP) as Global Frontier Project (NRF-2014M3A6B306370934). M.J.P., J.-W.R. and H.C.P. acknowledge financial support from the Institute for Basic Science in the Republic of Korea through the project IBS-R024-D1.

# Figures

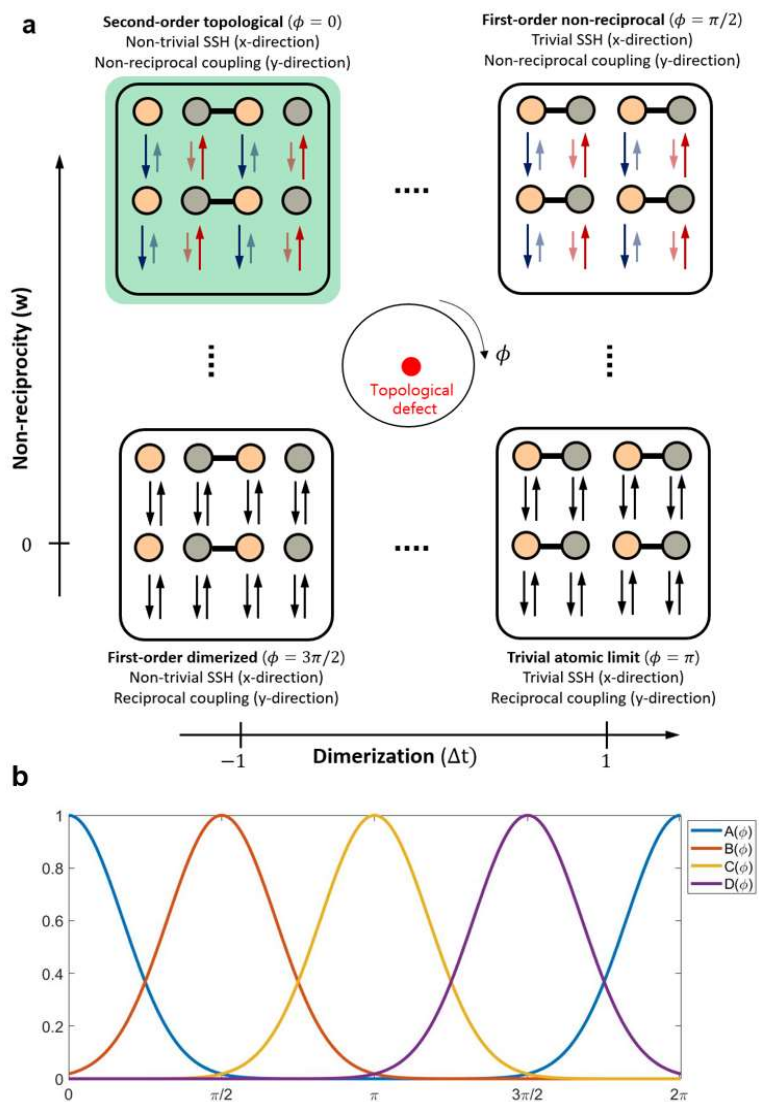

Extended data Fig. S1. **Adiabatic deformation scheme of the system a** Adiabatic deformation scheme of the Hamiltonian and the four distinct phases of the Hamiltonian configurations. Red and blue arrows indicate the nonreciprocal couplings along the *y*-direction. Black horizontal couplings represent the SSH-like coupling along the *x*-direction. For instance, the two-dimensional slice at $\phi = 0$ represents the topologically nontrivial Hamiltonian, while the slice at $\phi = \pi$ represents the vacuum Hamiltonian outside the circuit. The first-order nonreciprocal ($\phi = \pi/2$) and the first-order dimerized phases ($\phi = 3\pi/2$) correspond to the Hamiltonians where the *x*- and *y*-directional hoppings are, respectively trivialized. The skin states as a topological defect (drawn with a red dot) are guaranteed to exist by the presence of the Hopf bundle. **b** Plot of the Gaussian-shaped deformation parameters.